\documentclass[onecolumn]{whitepaper} 

\usepackage{blindtext}
\usepackage{wrapfig}
\usepackage[anythingbreaks]{breakurl}
\usepackage[rightcaption,raggedright]{sidecap}
   \sidecaptionvpos{figure}{t}
   \usepackage{caption}
   \DeclareCaptionLabelFormat{boldsmallperiod}{\bf{#1 #2.}}
\title{Connecting Solar and Stellar Flares/CMEs: Expanding Heliophysics\\ to Encompass Exoplanetary Space Weather}

\author[1]{Benjamin~J.~Lynch~\orcidlink{0000-0001-6886-855X}}
\author[2]{Brian~E.~Wood~\orcidlink{0000-0002-4998-0893}}
\author[3,4]{Meng~Jin~\orcidlink{0000-0002-9672-3873}}
\author[5]{Tibor~T\"{o}r\"{o}k~\orcidlink{0000-0003-3843-3242}}
\author[6]{Xudong~Sun~\orcidlink{0000-0003-4043-616X}}
\author[5]{Erika~Palmerio~\orcidlink{0000-0001-6590-3479}}
\author[7]{Rachel~A.~Osten~\orcidlink{0000-0001-5643-8421}}
\author[8]{Aline~A.~Vidotto~\orcidlink{0000-0001-5371-2675}}
\author[9]{Ofer~Cohen~\orcidlink{0000-0003-3721-0215}}
\author[10]{Juli\'{a}n~D.~Alvarado-G\'{o}mez~\orcidlink{0000-0001-5052-3473}}
\author[11]{Jeremy~J.~Drake~\orcidlink{0000-0002-0210-2276}}
\author[12,13]{Vladimir~S.~Airapetian~\orcidlink{0000-0003-4452-0588}}
\author[14,15]{Yuta~Notsu~\orcidlink{0000-0002-0412-0849}}
\author[16]{Astrid~Veronig~\orcidlink{0000-0003-2073-002X}}
\author[17]{Kosuke~Namekata~\orcidlink{0000-0002-1297-9485}}
\author[18]{R\'{e}ka~M.~Winslow~\orcidlink{0000-0002-9276-9487}}
\author[12]{Lan~K.~Jian~\orcidlink{0000-0002-6849-5527}}
\author[19]{Angelos~Vourlidas~\orcidlink{0000-0002-8164-5948}}
\author[18]{No\'{e}~Lugaz~\orcidlink{0000-0002-1890-6156}}
\author[18]{Nada~Al-Haddad~\orcidlink{0000-0002-0973-2027}}
\author[20]{Ward~B.~Manchester~\orcidlink{0000-0003-0472-9408}}
\author[18]{Camilla~Scolini~\orcidlink{0000-0002-5681-0526}}
\author[18]{Charles~J.~Farrugia~\orcidlink{0000-0001-8780-0673}}
\author[18]{Emma~E.~Davies~\orcidlink{0000-0001-9992-8471}}
\author[12]{Teresa~Nieves-Chinchilla~\orcidlink{0000-0003-0565-4890}}
\author[12,21]{Fernando~Carcaboso~\orcidlink{0000-0003-1758-6194}}
\author[1]{Christina~O.~Lee~\orcidlink{0000-0002-1604-3326}}
\author[12,22]{Tarik~M.~Salman~\orcidlink{0000-0001-6813-5671}}

\affil[1]{Space Sciences Laboratory, University of California--Berkeley, Berkeley, CA 94720, USA}
\affil[2]{Space Science Division, Naval Research Laboratory, Washington, DC 20375, USA}
\affil[3]{Lockheed Martin Solar and Astrophysics Laboratory, Palo Alto, CA 94304, USA}
\affil[4]{SETI Institute, Mountain View, CA 94043, USA}
\affil[5]{Predictive Science Inc., San Diego, CA 92121, USA}
\affil[6]{Institute for Astronomy, University of Hawai`i at M\={a}noa, Pukalani, HI 96768, USA}
\affil[7]{Space Telescope Science Institute, Baltimore, MD 21218, USA}
\affil[8]{Leiden Observatory, Leiden University, 2300 RA, Leiden, The Netherlands}
\affil[9]{Department of Physics and Applied Physics, University of Massachusetts--Lowell, Lowell, MA 01854, USA}
\affil[10]{Leibniz Institute for Astrophysics, 14482 Potsdam, Germany}
\affil[11]{Smithsonian Astrophysical Observatory, Cambridge, MA 02138, USA}
\affil[12]{Heliophysics Science Division, NASA Goddard Space Flight Center, Greenbelt, MD 20771, USA}
\affil[13]{American University, Washington, DC 20016 USA}
\affil[14]{Laboratory for Atmospheric and Space Physics, University of Colorado Boulder, Boulder, CO 80303, USA}
\affil[15]{National Solar Observatory, Boulder, CO 80303, USA}
\affil[16]{Institute of Physics, University of Graz, 8010 Graz, Austria}
\affil[17]{National Astronomical Observatory of Japan, Tokyo 181-8588, Japan}
\affil[18]{Space Science Center, University of New Hampshire, Durham, NH 03824, USA}
\affil[19]{Johns Hopkins University Applied Physics Laboratory, Laurel, MD 20723, USA}
\affil[20]{Department of Climate and Space Research, University of Michigan, Ann Arbor, MI 48109, USA}
\affil[21]{The Catholic University of America, Washington, DC 20064, USA}
\affil[22]{George Mason University, Fairfax, VA 22030, USA}

\runningtitle{Connecting Solar and Stellar Flares/CMEs}
\shortauthors{Lynch et al.}
\footer{White Paper submitted to the Heliophysics 2024--2033 Decadal Survey}

\begin{document}

\maketitle

\thispagestyle{firststyle}


\begin{abstract}


The aim of this white paper is to briefly summarize some of the outstanding gaps in the observations and modeling of stellar flares, CMEs, and exoplanetary space weather, and to discuss how the theoretical and computational tools and methods that have been developed in heliophysics can play a critical role in meeting these challenges. The maturity of data-inspired and data-constrained modeling of the Sun-to-Earth space weather chain provides a natural starting point for the development of new, multidisciplinary research and applications to other stars and their exoplanetary systems.
Here we present recommendations for future solar CME research to further advance stellar flare and CME studies. These recommendations will require institutional and funding agency support for both \emph{fundamental} research (e.g.\ theoretical considerations and idealized eruptive flare/CME numerical modeling) and \emph{applied} research (e.g.\ data inspired/constrained modeling and estimating exoplanetary space weather impacts). In short, we recommend continued and expanded support for:
(1.) Theoretical and numerical studies of CME initiation and low coronal evolution, including confinement of ``failed'' eruptions; 
(2.) Systematic analyses of Sun-as-a-star observations to develop and improve stellar CME detection techniques and alternatives;
(3.) Improvements in data-inspired and data-constrained MHD modeling of solar CMEs and their application to stellar systems; and
(4.) Encouraging comprehensive solar--stellar research collaborations and conferences through new interdisciplinary and multi-agency/division funding mechanisms.

\end{abstract}


\section{Introduction}

The aim of this white paper is to discuss the importance of both \emph{fundamental} and \emph{applied} coronal mass ejection (CME) research in the context of its increasing relevance to stellar astronomy. Interest in detecting and modeling CMEs on other stars has increased dramatically in recent years. The winds and CMEs of coronal stars like the Sun have always been of interest due to their role in shedding angular momentum, leading to observed declines in stellar rotation and activity with age \citep{vidotto21}. However, by far the primary driver of interest in stellar winds and CMEs these days relates to star--planet interactions and the ``space weather'' impacts on exoplanetary atmospheres \citep[][and references therein]{Airapetian2020}.

As of March 2022, there are over 5000 confirmed exoplanet discoveries \citep{brennan2022}. Most of the known exoplanets orbit very close to their parent stars, meaning they are potentially exposed to particularly high particle fluxes from stellar winds to CMEs, leading to much interest in the long-term effects this exposure has on the atmospheres of these planets. Absorption from material evaporating from planetary atmospheres has actually been detected in cases of transiting exoplanets, indicating the importance of this process \citep{vidal-madjar03, lecavelier10, ekenback10, kislyakova14, bourrier16, schneiter16}. In our own solar system, solar wind and CME exposure may have significantly affected planetary atmospheric evolution, with Mars being a particularly interesting case \citep{jakosky18}.

Solar and stellar flares---sudden explosive releases of energy in the solar/stellar atmosphere across a wide range of electromagnetic wavelengths---occur due to the rapid release of free magnetic energy stored in the sheared and/or twisted strong fields typically associated with sunspots and active regions \citep{Forbes2000a, Fletcher2011, Shibata2011, Kazachenko2012}. The onset and evolution of solar and stellar flares are intimately coupled to magnetic reconnection processes \citep{Klimchuk2001, Green2018}. The long-standing CSHKP model \citep{Carmichael1964, Sturrock1966, Hirayama1974, Kopp1976} for eruptive solar flares explains many of their observational properties \citep[e.g.][]{Janvier2015, Torok2018, Lynch2021}. 
Large flares are often accompanied by CMEs \citep{Andrews2003, Gopalswamy2005} and CMEs are largely responsible for the most geoeffective space-weather impacts at Earth and other solar system bodies \citep[][]{ZhangJ2021}. 


Magnetohydrodynamic (MHD) modeling of stellar CMEs and their interactions with exoplanets began not long after exoplanets were discovered \citep{khodachenko07, lammer07}. Many of these models utilize the same codes used to model solar CME propagation in the heliosphere and interaction with Earth's magnetosphere \citep{cohen11, garraffo16, cherenkov17, lynch19, hazra22}.
In this white paper, we present a brief summary of the applications of state-of-the-art MHD models used in heliophysics to stellar magnetic environments (Section~\ref{sec2}) and their exoplanetary systems (Section~\ref{sec3}) in order to discuss the current observational and modeling limitations. In Section~\ref{sec4}, we conclude with some recommendations for future research strategies in order to advance our understanding of the solar--stellar connection.

\section{Current Theoretical and Observational Ambiguities and/or Discrepancies} \label{sec2}

\subsection{Measurements of Stellar Magnetic Fields (and their Limitations)} \label{sec2.1}

\begin{SCfigure}[36][t]
	\includegraphics[width=0.65\textwidth]{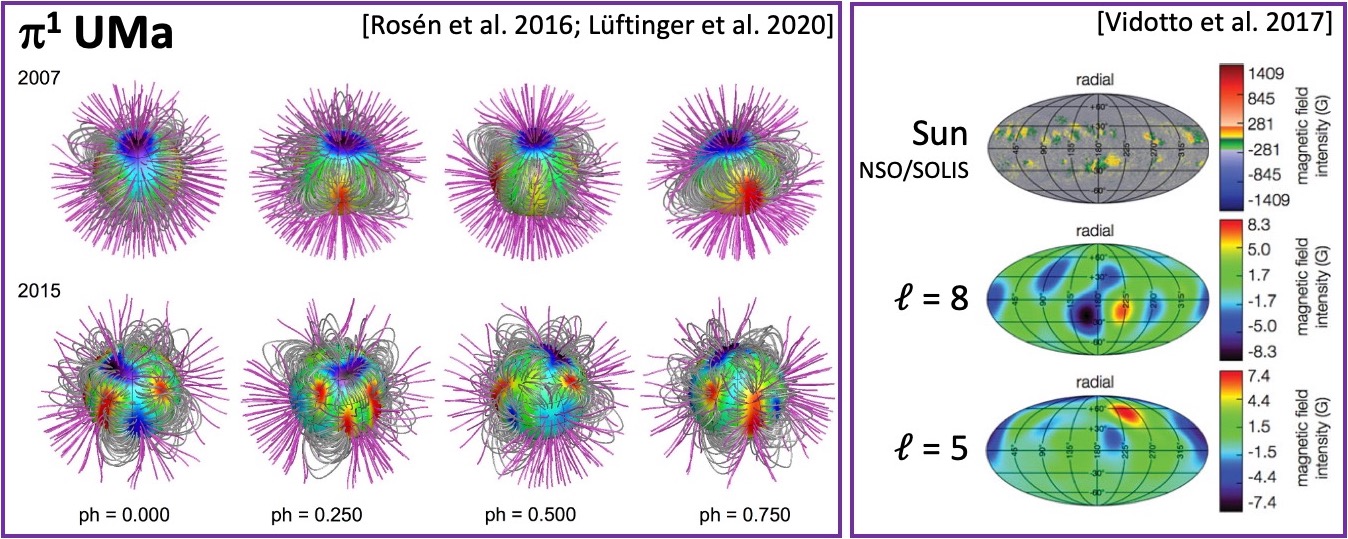}
	\caption{Left panel: ZDI stellar magnetograms of $\pi^1$~UMa obtained in 2007 and 2015 \citep[from][]{Rosen2016,Lueftinger2020}. Right panel: Comparison between solar $B_r$ magnetic field observations from NSO/SOLIS (top) with two low-degree PFSS reconstructions \citep[from][]{Vidotto2017}.}
	\label{fig:zdi}
\end{SCfigure}

For decades, the magnetic fields of massive, early-type stars were analyzed assuming simple dipole or dipole-plus-quadrupole magnetic field geometries. For late-type active stars, the development of Zeeman--Doppler Imaging \citep[ZDI;][]{Dontai1997MNRAS, Piskunov2002, Kochukhov2016} and its inversion techniques has made it possible to resolve---at least on the largest scales---surface magnetic field distributions that can be considerably more complex and track their long-term evolution, e.g. the polarity reversals associated with stellar activity cycles \citep{Lueftinger2015, Kochukhov2016}. 
Temperature or abundance structures on the surface of stars can also be reconstructed by inverting time series of high-resolution spectropolarimetric data in the Stokes I, V, Q, and U profiles \citep[e.g.][]{Lueftinger2010b,Lueftinger2010a}. 
The left panel of Figure~\ref{fig:zdi} \citep[adapted from][]{Rosen2016,Lueftinger2020}, shows the ZDI magnetic field structure obtained for $\pi^1$~UMa during 2007 (top row, showing a relatively simple ``solar minimum'' configuration) and 2015 (bottom row, showing a more complex ``solar maximum'' configuration).
The availability of stellar magnetic field maps has significantly advanced our capacity for sophisticated numerical modeling of stellar coronae, winds, and star–planet interactions \citep{cohen11, Vidotto2011, doNascimento2016, garraffo16, alvarado-gomez18}.

\emph{\textbf{A current limitation of the ZDI technique is illustrated in the right panel of Figure~\ref{fig:zdi}}} \citep[adapted from][]{Vidotto2017}. The top row shows the vector magnetic field components obtained via NSO/SOLIS \citep{Pevtsov2010,Bertello2013} whereas the middle and bottom rows show low-degree spherical harmonic representations ($\ell = 8$ and $\ell = 5$). In addition to lacking the overall spatial resolution required to resolve the strong-field active region flux systems observed on the Sun, the maximum field strengths are significantly underestimated (by roughly 2 orders of magnitude).
In the young solar analog $\kappa^1$~Cet, ZDI synoptic magnetograms have maximum surface field strengths that are on the order of $\sim$20~G.
However, measurements of unsigned magnetic flux from Zeeman broadening of the unpolarized Stokes~I spectra suggest that $\kappa^{1}$~Cet has a disk-averaged, magnetic field magnitude of $\langle \, fB \, \rangle \sim 500$~G \citep[e.g.][]{Saar1992, Kochukhov2020}. This is the result of an unresolved magnetic flux in Stokes~V (circularly polarized) observations due to an effective ``cancellation'' of the oppositely signed magnetic flux of starspots within pixel resolution via suppression of the Zeeman effect in dark regions \citep{Kochukhov2016}. Thus, up to 90--95\% of the magnetic flux is concentrated in small magnetic structures represented by stellar active regions/starspots that remain unresolved with current ZDI techniques \cite[e.g.][]{Reiners2009, See2019, Kochukhov2016, Kochukhov2020}.

In general, the energy of solar/stellar flares ($E_{\rm flare}$) will be proportional to the free magnetic energy ($E_M$) stored in energized coronal magnetic field structures and thus can be written as
$E_{\rm flare} \approx f_c  E_{\rm M} = f_c \left( 8\pi \right)^{-1} \left( B_{AR} \right)^2 \left( A_{AR} \right)^{3/2}$
%
%
\citep{Shibata2013,Maehara2015} where $f_c$ is a coefficient ($\le 1$) describing the energy partition \citep{Emslie2012}. To obtain the stellar flare/CME energies corresponding to so-called stellar ``superflares'' ($10^{\, \gtrsim 34}$~erg) in numerical MHD models, one must construct sufficiently strong magnetic fields over a large enough area. 
 
\begin{figure}[t]
	\centering
	\includegraphics[width=0.90\textwidth,height=2.6in]{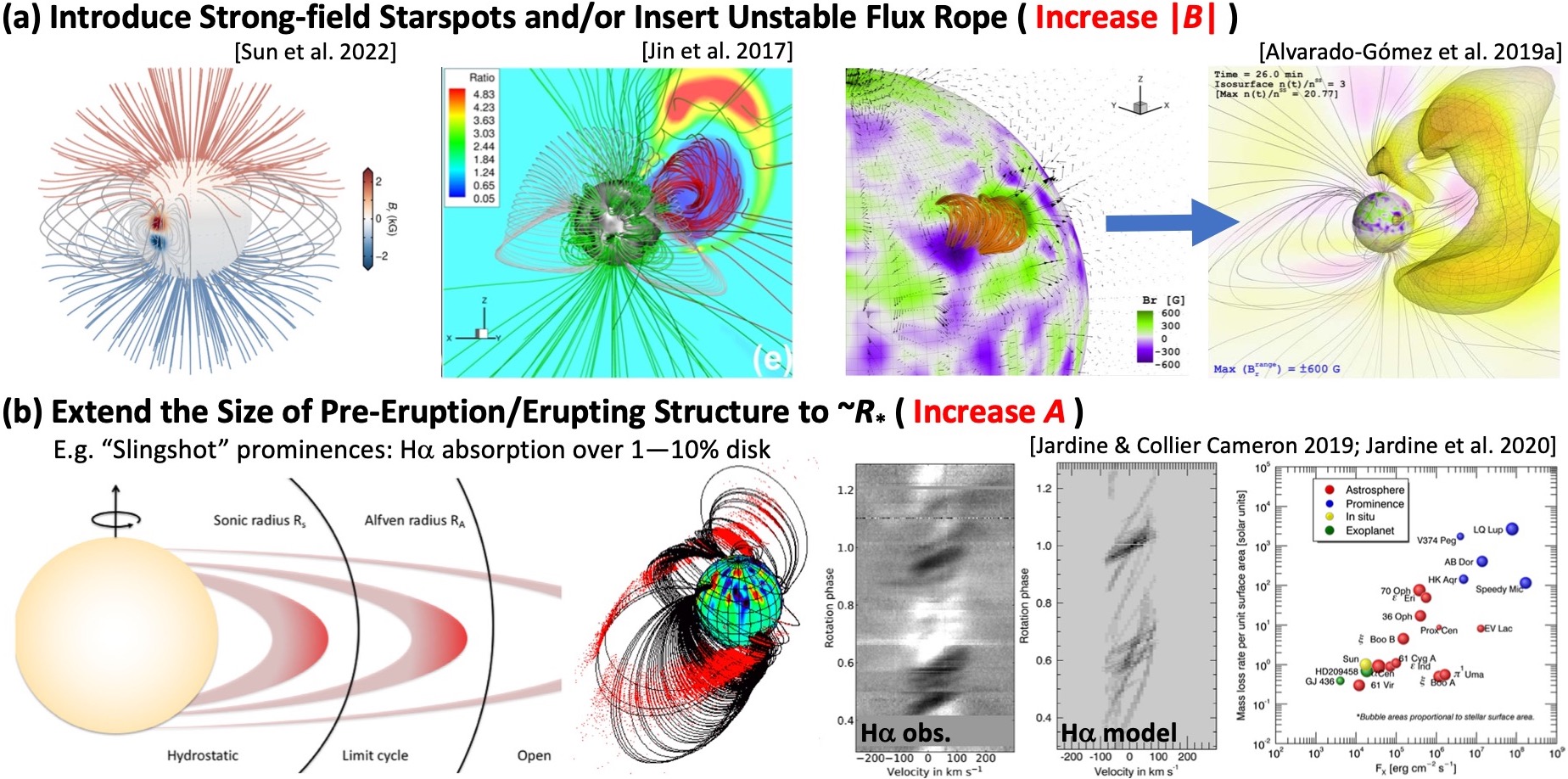}
	\caption{Two approaches to compensate for unresolved starspots/active region flux in order to obtain stellar superflare/CME energies of 10$^{\, \gtrsim 34}$~erg consistent with observations. (a) Increasing the effective magnetic field strength of the CME source region through the addition of strong-field starspot active regions \citep{Sun2022}, the insertion of a highly energized, unstable flux rope \citep{Jin2017}, or some combination of the two \citep{alvarado-gomez19a}. (b) Increasing the effective area (size) of the CME source region through modeling large-scale, pre-eruption structures such as ``slingshot prominences'' which may regularly span the entire disk \citep{Jardine2019,Jardine2020}.}
	\label{fig:cme}
\end{figure}

\emph{\textbf{How do we deal with unresolved stellar active region flux?}} 
Figure~\ref{fig:cme} summarizes two (complementary) approaches for increasing $E_M$. The first approach is shown in Figure~\ref{fig:cme}a, which illustrates examples of modeling large $|B|$ via the introduction of strong-field star spots \citep{Sun2022}, inserting an unstable magnetic flux rope \citep{Jin2017}, and what is effectively a combination of the two \citep{alvarado-gomez19a}.
To constrain the field magnitude $B$, one can estimate the starspot/active region areas (filling-factor $f$ in the $\langle f B \rangle$ measurement), via transit observations where large starspot/active regions are usually identified as rotationally-modulated ``dips'' from dark regions on the stellar disk \citep{Namekata2019}. Again, in the example of $\kappa{^1}$~Cet, \citet{Rucinski2004} and \citet{Walker2007} determined the light curve variations were consistent with two large starspots in 2003 (with areas of 1.4\% and 3.6\% of the stellar disk), three main starspots in 2004 (with areas of 1.9\%, 5.3\%, and 9\% of the disk), and two spots in 2005 (2.2\% and 2.9\% of the stellar disk). These are at least a factor of 10 larger than the largest observed solar active region complexes \citep{Hoge1947}. Future stellar observations could provide valuable information about smaller starspots/bipolar structures in optical and Far UV bands during their transit, as in the Sun-as-a-star study of \citet{Toriumi2020}.

\begin{SCfigure}[36][t]
	\includegraphics[width=0.75\textwidth,trim=0 0 8.50in 0,clip]{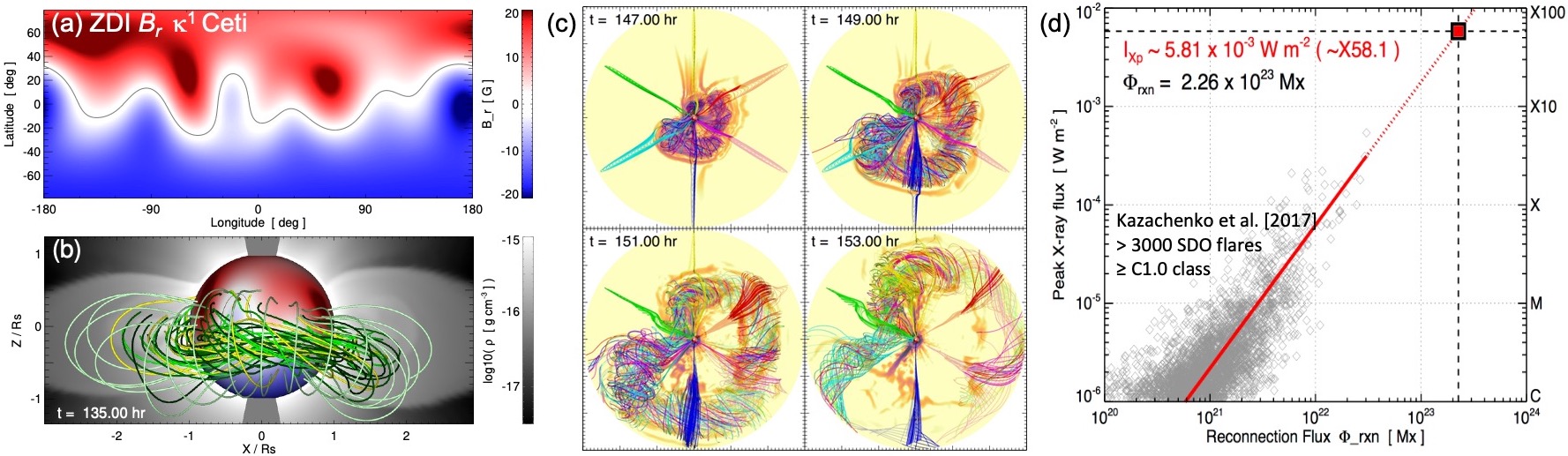}
	\caption{Overview of the \citet{lynch19} simulation of a Carrington-scale, X58 superflare and CME from the young solar analog $\kappa{^1}$~Cet. (a) ZDI magnetogram. (b) Energized, pre-eruption field structure. (c) CME evolution through 30$R_\odot$.}
	\label{fig:lynch}
\end{SCfigure}

The second approach is shown in Figure~\ref{fig:cme}b which presents a simple model for a pre-eruption, (potentially unstable) ``slingshot prominence'' structure that was estimated to cover a significant fraction of the stellar surface, i.e.\ large $A$, in order to explain observed H$\alpha$ absorption features \citep{Jardine2019,Jardine2020}.
The Figure~\ref{fig:cme}b strategy of utilizing the largest possible source region area was employed by \citet{lynch19} to energize and erupt a 360$^{\circ}$-wide streamer blowout CME.
Figure~\ref{fig:lynch} shows an overview of the \citet{lynch19} MHD simulation of a Carrington-scale eruptive flare and CME modeled with the ZDI synoptic magnetogram calculated by \citet{Rosen2016} for $\kappa{^1}$~Cet in August of 2012. 
Figure~\ref{fig:lynch}a,b show the stellar $B_r$ distribution and the global-scale, pre-eruption prominence-like field structure while Figure~\ref{fig:lynch}c shows a series of snapshots of the eruption in the ecliptic plane.

\begin{itemize}[leftmargin=0pt,itemsep=0pt]
\item[\textcolor{magenta}{$\bullet$}] \textcolor{magenta}{Future modeling of stellar flare/CMEs will likely require one or both strategies to address the observational uncertainties in global and local magnetic field configurations of the flaring/CME source regions and should explore the eruption parameter space and resulting energy partition.}
\end{itemize}

\subsection{Measurements of Stellar Winds and Stellar CMEs (and their Limitations)}  \label{sec2.2}

A fundamental difficulty with stellar wind research is that it is extremely hard to detect any component of coronal stellar winds, whether quiescent wind or transient CMEs. The most successful technique for studying stellar winds so far is by detecting hydrogen Lyman-$\alpha$ absorption from interaction regions between the winds and the surrounding interstellar medium, i.e.\ astrospheric absorption \citep{linsky96, wood05}. However, even this technique has so far led to only 22 wind detections/measurements, and 7 useful upper limits \citep{wood21}.  Furthermore, this diagnostic is measuring the average wind ram pressure over long timescales, typically years to decades depending on the size of the astrosphere. Thus, it is unknown whether the observed stellar winds are dominated by quiescent wind or CMEs.

Candidate stellar CME detections are typically via observations not commonly used in solar CME research, meaning that it is not entirely certain the same phenomenon is being seen. 
That being said, significant progress has been made in the characterization and interpretation of stellar CME signatures by utilizing the results of multi-wavelength Sun-as-a-star analyses of large CME events \citep[e.g.\ ][]{Leitzinger2022, Namekata2022, namekata21, Xu2022}.

A number of stellar CME claims originate from detection of blueshifted H$\alpha$ emission or absorption after stellar flares \citep{leitzinger20, muheki20, odert20, namekata21}. On the Sun, such observations would be called signatures of prominence/filament eruptions.  While there are certainly cases where prominence material ends up incorporated into a CME that escapes the Sun \citep[e.g.][]{wood17,Lepri2014IAU,Lepri2021}, this is not always the case, so an H$\alpha$ signature by itself would not necessarily be considered a CME detection. Blueshifted coronal lines observed after stellar flares have also been observed \citep{argiroffi19,namekata21}.

One solar CME detection technique that does have potential applicability to how stars are observed is coronal dimming, demonstrated using
full-disk SDO/EVE observations of low temperature coronal lines like
Fe~IX $\lambda$171 ($\log_{10} T/K \sim 5.8$) [\citealt{mason16}; see also \citealt{Harra2016}]. There are post-flare
coronal dimmings that have been observed on stars, which have been interpreted as possible CMEs \citep{veronig21,loyd2022}. 
However, there is not a one-to-one correspondence between dimmings and the loss of coronal material via eruption, e.g.\ confined eruptions can also show dimming profiles at coronal temperatures \citep{alvarado-gomez19a}. 

Type II radio bursts are another promising stellar CME detection technique analogous to how CMEs (specifically, CME-driven shocks) are observed on the Sun. Such observations would have the added benefit of indicating the CME speed through the rate of change in radio frequency \citep{reiner07, liu13}.  Unfortunately, attempts to detect Type II bursts from frequently flaring M dwarfs, have so far proved unsuccessful \citep{crosley18a, crosley18b, villadsen19}. 
While the solar CME--Type II association rate is only 4\% overall \citep{Bilenko2018}, it is much higher for the most energetic CME events.
The stellar Type II nondetections call into question the existence of fast, massive CMEs that are assumed to accompany the extremely energetic flares from M dwarf stars.
The density jump of the CME-driven shock in the \citet{lynch19} simulation ``predicted'' electron plasma frequencies below the ionospheric cutoff of $\lesssim$20~MHz and the suppression/slow-down of stellar CMEs combined with the higher Alfv\'{e}n speed profiles for active stars will result in a similar outcome \citep{alvarado-gomez20b}, making ground-based detection extremely difficult.

\begin{itemize}[leftmargin=0pt,itemsep=0pt]
\item[\textcolor{magenta}{$\bullet$}] \textcolor{magenta}{Simultaneous multi-wavelength observations of potential stellar CME signatures (e.g.\ H$\alpha$, X-ray and UV dimmings, radio bursts) are expected to makes significant progress toward resolving at least some of the current observational ambiguities, especially with complementary Sun-as-a-star analyses of solar flares/filament eruptions/CMEs in H$\alpha$ and EUV wavelengths. Both idealized and ``data-constrained'' modeling of stellar winds and flares/CMEs should aim to generate synthetic observables at different wavelengths for observational guidance and to quantify detectability thresholds.}
\end{itemize}

\subsection{Do Solar--Stellar ``Scaling Laws'' Break Down?} \label{sec2.3}
  
\begin{SCfigure}[36][t]
	\includegraphics[width=0.78\textwidth,height=2.35in]{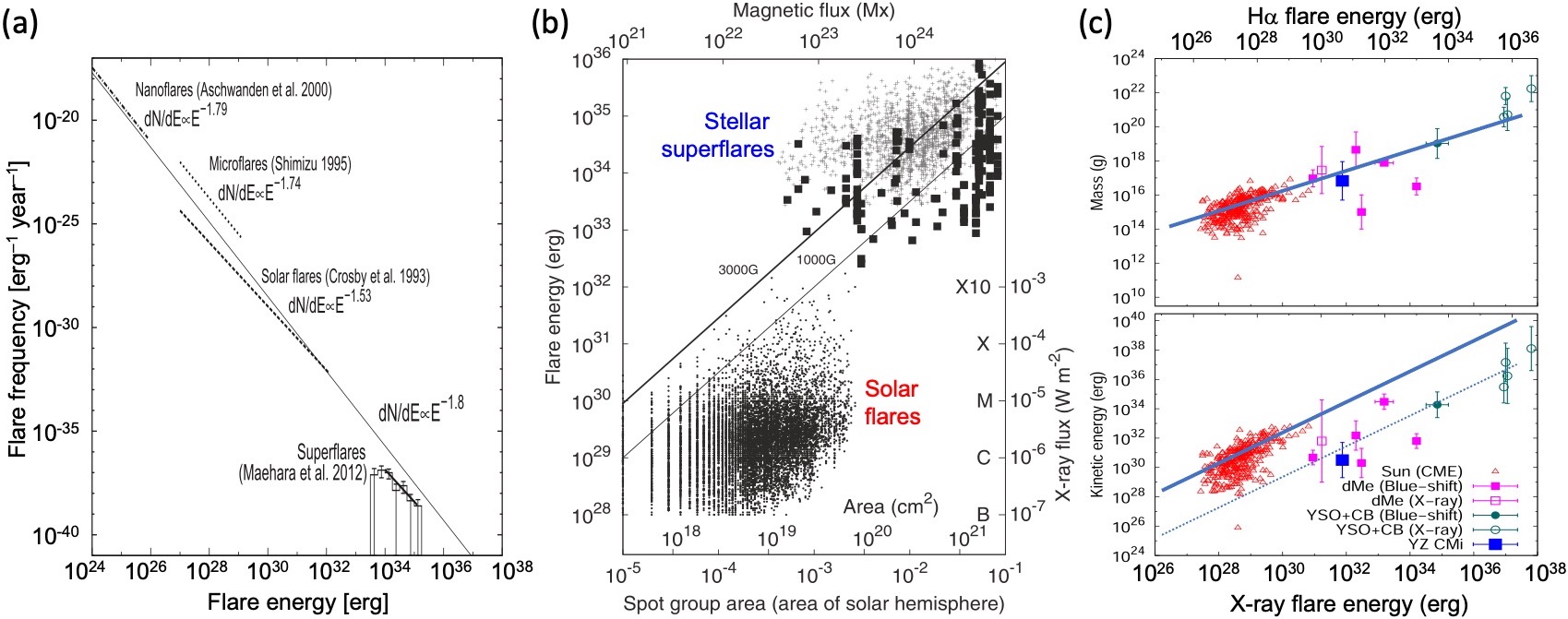}
	\caption{Solar--stellar scaling laws. (a) Flare frequency~vs.~energy \citep{Shibata2013}. (b) Flare energy~vs.~active region area \citep{Maehara2015}. (c)  CME mass and kinetic energy~vs.~X-ray and H$\alpha$ flare energy \citep{Maehara2021}. }
	\label{fig:scaling}
\end{SCfigure}

For the Sun, a strong correlation is found between flare strength as quantified by X-ray luminosity and CME mass \citep[e.g.][]{aarnio11}. Given that we have very limited observational knowledge about the nature of CMEs emanating from other stars, it is natural to apply solar flare/CME relations to active stars that flare more frequently and energetically, in order to estimate what CMEs might contribute to the stellar winds of these stars \citep{moschou19}.

Figure~\ref{fig:scaling} demonstrates a number of solar--stellar scaling laws. 
Figure~\ref{fig:scaling}a, adapted from \citet{Shibata2013}, plots the power-law distributions obtained for flare occurrence frequency as a function of energy, spanning $\sim$10 orders of magnitude from nanoflares \citep{Aschwanden2000}, to microflares \citep{Shimizu1995}, ``standard'' solar flares \citep{Crosby1993}, and stellar superflares \citep{Maehara2012}. The power-law $dN/dE \propto E^{-1.8}$ is shown as the thin solid line and appears reasonably consistent with the observed occurrence frequencies over the entire flare energy range. 
Figure~\ref{fig:scaling}b, adapted from \citet{Maehara2015}, shows flare energy vs.\ solar/stellar spot area (and associated unsigned magnetic flux). The grouping of solar flares and stellar superflares are labeled accordingly and the thick (thin) solid line corresponds to the $E_{\rm flare}$ expression of Section~\ref{sec2.1} as a function of $A_{AR}$ assuming $f_c=0.10$ and constant $B_{AR}$ values of 1~kG (3~kG), respectively.

Figure~\ref{fig:scaling}c, adapted from \citet{Maehara2021}, shows the CME mass ($M_{\rm cme}$; upper panel) and kinetic energy ($E_K$; lower panel) estimates vs. flare (X-ray) energy (or equivalently, the flare H$\alpha$ energy as a simple rescaling of the X-ray energy). The set of solar flare--CME points \citep[from][]{Yashiro2009} are shown as red triangles, candidate stellar eruptive flare/CME detections as the blue, magenta, and teal points, along with the power-law fits obtained by \citet{drake13}: $M_{\rm cme} \propto E_{\rm X-ray}^{0.59\pm0.02}$ and $E_K \propto E_{\rm X-ray}^{1.05\pm0.03}$ (blue lines). It is interesting to note that the $M_{\rm cme}$ scaling appears consistent between the solar and stellar cases whereas $E_K$ estimated for the stellar events are comfortably below the extrapolated solar scaling law by 2--3 orders of magnitude. This may be a result of simply underestimating the stellar CME values because the prominence material (as the source of H$\alpha$ absorption/emission) is merely a subset of the larger CME erupting structure, or that these potential stellar CME detections are representative of magnetic environments that impede traditional ``solar-like'' CME eruptions and their evolution through extended stellar coronae \citep[e.g.\ as in][]{alvarado-gomez18}.

Merely extrapolating to stellar regimes from solar data for truly active stars invariably leads to conclusions that such stars should have winds hundreds or thousands of times stronger than the solar wind simply due to CMEs alone [e.g. \citealt{drake13, odert17}; Figure~\ref{fig:scaling}c]. Such conclusions not only conflict with the Type II radio burst nondetections, but they also conflict with a failure to detect such strong winds using the aforementioned astrospheric Lyman-$\alpha$ absorption technique \citep{wood21}. For example, the astrospheric measurements suggest mass-loss rates of only 1 and 30 times the solar mass-loss rate for the notorious M dwarf flare stars EV~Lac and YZ~CMi, respectively.
Clearly, the strong connection between flares and fast, massive CMEs on the Sun cannot extend to flare stars like EV Lac and YZ CMi. For such stars, CMEs must be far less common or far less massive than one might expect, given the frequent flaring. 

The Sun itself may provide clues for what is happening on active flare stars, as there are many cases of strong flares with no associated CME. A well-studied example is the  series of X-class flares from active region NOAA 12192, which was highly flare-productive, particularly in 2014 October. However, almost none of the flares from AR 12192 had associated CMEs \citep{sun15, thalmann15}. On the Sun, this is unusual, but on active stars perhaps this is the norm. Strong magnetic fields overlying an active region can inhibit CME eruption. Numerical simulations of CMEs on active stars made in recent years include models of such confined eruptions \citep{alvarado-gomez18, alvarado-gomez19a, alvarado-gomez19b, alvarado-gomez20b}, and \citet{Sun2022} performed a theoretical study of the susceptibility of stellar active regions to the torus instability for CME initiation \citep{Kliem2006}, which explores why active stars may be less prone to CMEs than generally supposed. 

\begin{itemize}[leftmargin=0pt,itemsep=0pt]
\item[\textcolor{magenta}{$\bullet$}] \textcolor{magenta}{We strongly encourage research designed to explore the physical conditions leading to both successful and ``confined'' eruptions in order to understand the regimes where the semi-empirical solar--stellar scaling laws appear to work and those where they do not.}
\end{itemize}

\section{Stellar Space Weather Impacts on Exoplanetary Atmospheres}
\label{sec3}

\begin{SCfigure}[36][t]
	
	\includegraphics[width=0.73\textwidth]{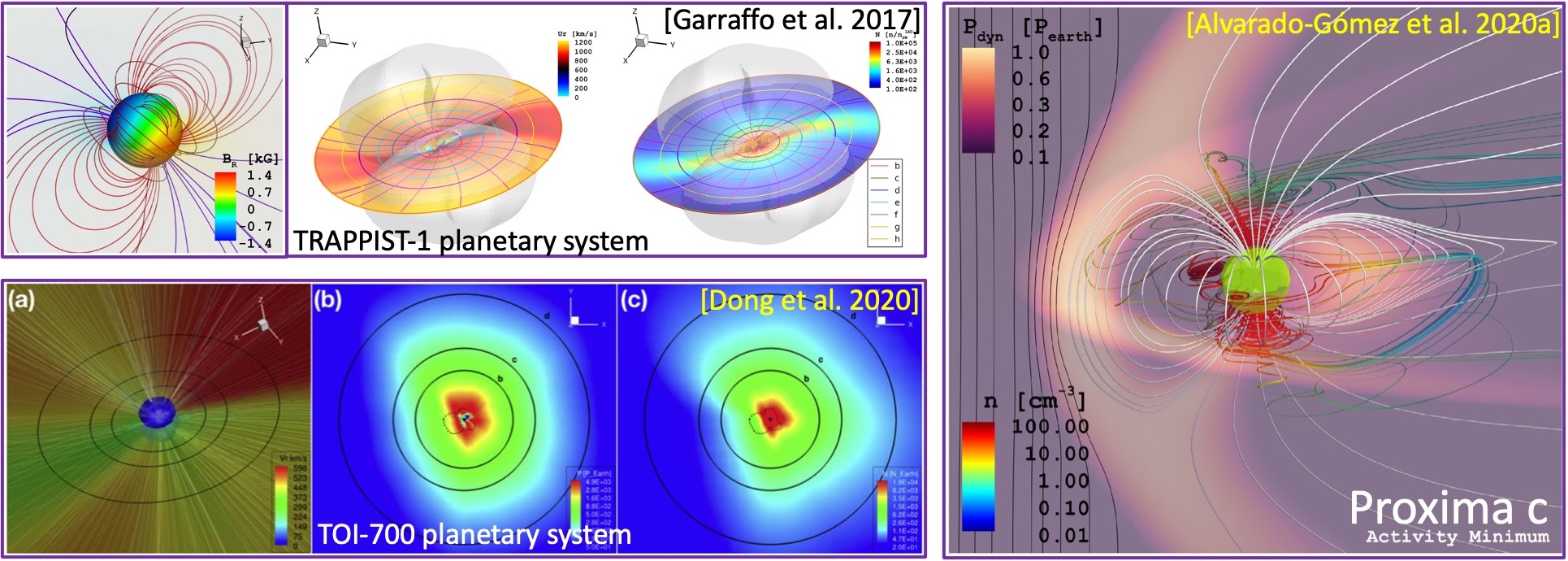}
	\caption{Modeling stellar winds and star--planet interactions: TRAPPIST-1 system \citep[upper left;][]{Garraffo2017}, TOI-700 system \citep[lower left;][]{Dong2020}, and for Proxima~c \citep[right;][]{alvarado-gomez20a}.}
	\label{fig:spi}
\end{SCfigure}

One important application of stellar wind measurements is to better understand the environment of exoplanets around cool stars. The M dwarfs are of particular interest due in part to the abundance of such stars in the galaxy. Also, the habitable zones of the intrinsically faint M dwarfs are much closer to the stars than for earlier type stars like the Sun, so planets in such locations will potentially be exposed to much higher particle fluxes from stellar winds. Assessments of the potential impact of this wind exposure on planets in M dwarf systems have been underway for some time [e.g. \citealt{vidotto13, garraffo16, dong17, alvarado-gomez19b, alvarado-gomez20a}]. Figure~\ref{fig:spi} presents three recent examples of star-to-planet modeling to explore the impact of steady-state stellar winds on exoplanets in the TRAPPIST-1 \citep[upper left;][]{Garraffo2017}, TOI-700 \citep[lower left;][]{Dong2020}, and Proxima Centari \citep[right;][]{alvarado-gomez20a} systems.

Whether M dwarf habitable zone planets are truly habitable is in part tied to the question of whether intense exposure of such stars to stellar flares, CMEs, and energetic particles would make habitability impossible [e.g. \citealt{khodachenko07, Yamashiki2019, Hu2022}]. If fast, massive CMEs from frequently flaring M dwarfs are less common than generally thought, perhaps CME exposure is not as big a factor for habitability as often supposed.  Furthermore, in the solar case, the most damaging energetic particles originate from CME shocks rather than flares, if fast CMEs are less common than generally thought, perhaps energetic particle fluxes are also lower \citep{fraschetti19}. Exoplanets in M dwarf habitable zones will certainly be exposed to high X-ray fluxes, both from quiescent  coronal emission and flares, but it remains an open question whether solar models of coronal heating and wind acceleration can be applied to M dwarfs and how significant stellar CMEs and energetic particle fluxes are to long-term atmospheric evolution, and ultimately, to exoplanet habitability.

\begin{itemize}[leftmargin=0pt,itemsep=0pt]
\item[\textcolor{magenta}{$\bullet$}] \textcolor{magenta}{Future application of heliophysics modeling tools to star-to-planet systems should aspire to be as ``data-constrained'' as possible (to limit the wind and flare/CME model parameter spaces) and include forward modeling of synthetic observable quantities, i.e.~spectral signatures of atmospheric composition, chemistry and evolution/loss, auroral emission, etc.}
\end{itemize}













\section{Recommendations for Future Research Directions}
\label{sec4}


\begin{itemize}[leftmargin=3ex,itemsep=3pt]

\item[1.] \textcolor{blue}{Theoretical and numerical studies of CME initiation and low coronal evolution, including confinement of ``failed'' eruptions.}  Existing measurements of active stars suggest that most(?) of these stellar flares may represent confined eruptions without an accompanying ``solar-like'' CME.  Future research should characterize solar active region sources of confined eruptions, quantify the confined/eruptive thresholds, and investigate the implications for stellar magnetic field configurations that inhibit or facilitate CMEs from stellar superflares.

\item[2.] \textcolor{blue}{Systematic analyses of Sun-as-a-star observations to develop and improve stellar CME detection techniques and alternatives}, including but not limited to: disk-integrated brightening and dimming in EUV and X-ray spectral lines/wavelength ranges; quantitative forward modeling of flare, CME, and CME-driven shock radio emission; data analysis and forward modeling of eruption-induced Doppler shifts in H$\alpha$ and UV lines.

\item[3.] \textcolor{blue}{Improvements in data-inspired and data-constrained MHD modeling of solar CMEs and their application to stellar systems.} Develop holistic, full star-to-planet system modeling (or sequence of models) to characterize the range of exoplanetary space weather star--planet interactions and their impact on exoplanetary atmospheres over evolutionary timescales.

\item[4.] \textcolor{blue}{Encourage comprehensive solar--stellar research collaborations and conferences through new interdisciplinary and multi-agency/division funding mechanisms.} The multidisciplinary nature of this type of solar--stellar research is likely to require coordinated ``Centers of Excellence'' organizational/institutional support somewhat analogous to the NASA Astrobiology Institutes, the Heliophysics DRIVE centers, or the joint NSF--NASA funding structures that have supported multi-institution Space Weather research networks. Smaller focused efforts should also be supported through, e.g., future NASA LWS FST topics, expanded XRP funding, and other opportunities with solar/stellar overlap such as NSF AAG/AGS programs.  

\end{itemize}

\newpage

\bibliographystyle{wp_bibstyle}
\bibliography{bibliography}

\end{document}